\documentclass[english,amsmath,amssymb,aps,prb,showpacs,twocolumn]{revtex4-1}
\pdfoutput=1
\usepackage[T1]{fontenc}
\usepackage{amsmath}
\usepackage{amssymb}
\usepackage{graphicx}
\usepackage{esint}
\usepackage{caption}
\usepackage{ulem}
\usepackage{subcaption}
\usepackage{color}
\usepackage{babel}


\begin{document}

\title{Controlling Spin-Polarization in Graphene by Cloaking Magnetic and Spin-Orbit Scatterers}
\author{Diego Oliver$^1$ and Tatiana G. Rappoport$^1$}
\affiliation{$^1$Instituto de F\'isica, Universidade Federal do Rio de Janeiro, Caixa Postal 68528, 21941-972 Rio de Janeiro RJ, Brazil}

\begin{abstract} 

We consider spin-dependent  scatterers with large scattering cross-sections in graphene -a Zeeman-like and an intrinsic spin-orbit coupling impurity- and show that a gated ring around them can be engineered to produce an efficient control of the spin dependent transport, like current spin polarization and spin Hall angle. Our analysis is based on a spin-dependent partial-waves expansion of the electronic wave-functions in the continuum approximation, described by the Dirac equation.

\end{abstract}

\maketitle

\section{Introduction}
\label{sec1}
Graphene is a non-magnetic material with very weak spin-orbit coupling (SOC). This characteristic, associated with high electronic mobilities\cite{CastroNeto2009a, DasSarma2011} and spin diffusion lengths of several micrometers at room temperature \cite{Tombros2007, Han2010, Zomer2012, Yang2011, Han2011, Dlubak2012} makes it a promising candidate as spin conductor. Since its discovery,  a large number of theoretical studies proposed to introduce spin-dependent properties in graphene, either by inducing magnetism~\cite{Peres2005,Yazyev2007, Uchoa2008, Boukhvalov2008,Rappoport2009} or spin-orbit coupling~\cite{Weeks2011,Zhang2012, Ferreira2014,Milletari2016,Cazalilla2016,Garcia2016}. In recent years,  there has been a significant progress in engineering those properties. Several experimental studies observed magnetic moments in graphene as a result of vacancies, adatoms and molecular doping \cite{Hong2012, Nair2012, Nair2013,McCreary2012,Giesbers2013, Science2016}. More recently,  proximity effect with a magnetic insulator led to ferrromagnetism and spin polarized carriers~\cite{Wang2015}. Spin orbit coupling was also induced successfully in graphene doped with adatoms and clusters~\cite{Balakrishnan2013,Balakrishnan2014} and through proximity effect with other materials with strong SOC~\cite{Avsar2014,Calleja2015, Morpurgo2015}. The enhancement of  SOC  led to the observation of the spin Hall effect (SHE)~\cite{Balakrishnan2013,Balakrishnan2014}, while spin pumping experiments opened the possibility to transform spin to electrical currents  in graphene in an efficient way~\cite{Mendes2015}.

The analogies between light and electrons are known to originate a large number of effects observed in condensed matter and in optics, as for example,  the Fano resonances \cite{Miroshnichenko2010} and the Anderson localization \cite{Wiersma2013}. These resemblances have been exploited to produce electronic devices in analogy with photonic ones, such as beam-splitters, wave-guides, Faby-Perot interferometers in ballistic graphene \cite{Zhang2009, Rickhaus2013} and graphene based devices analogues to optical applications of the metamaterials \cite{Cheianov2007, Silveirinha2013}. The cloaking mechanism, in the first approach, refers to the process of invisibility present in optics. The main idea is to use a cloak with specific characteristics to envelop the target that we want to camouflage. Advances in the aspect of the metamaterials \cite{Zheludev2012} allow the production of electromagnetic cloaks to achieve the invisibility. Between the several techniques used in the cloaking procedure, we can highlight the coordinate-transformation method \cite{Leonhardt2006, Schurig2006, Pendry2006} and the scattering cancellation technique \cite{Alu2005, Alu2008, Edwards2009, Filonov2012, Chen2012, KortKamp2013, Rainwater2012, Nicorovici1994}. Bringing the cloaking mechanism into the idea of electronic devices, we can use a method based on the partial waves expansion \cite{Liao2012}, which was applied latter to graphene \cite{Liao2013,Oliver2015}, that demonstrates that nanoparticles adsorbed in a semiconductor with a size comparable to the wave length of the electron incident beam can be "invisible" to the incoming wave \cite{Liao2012}.

\begin{figure}[h]
\includegraphics[clip,width=0.9\columnwidth]{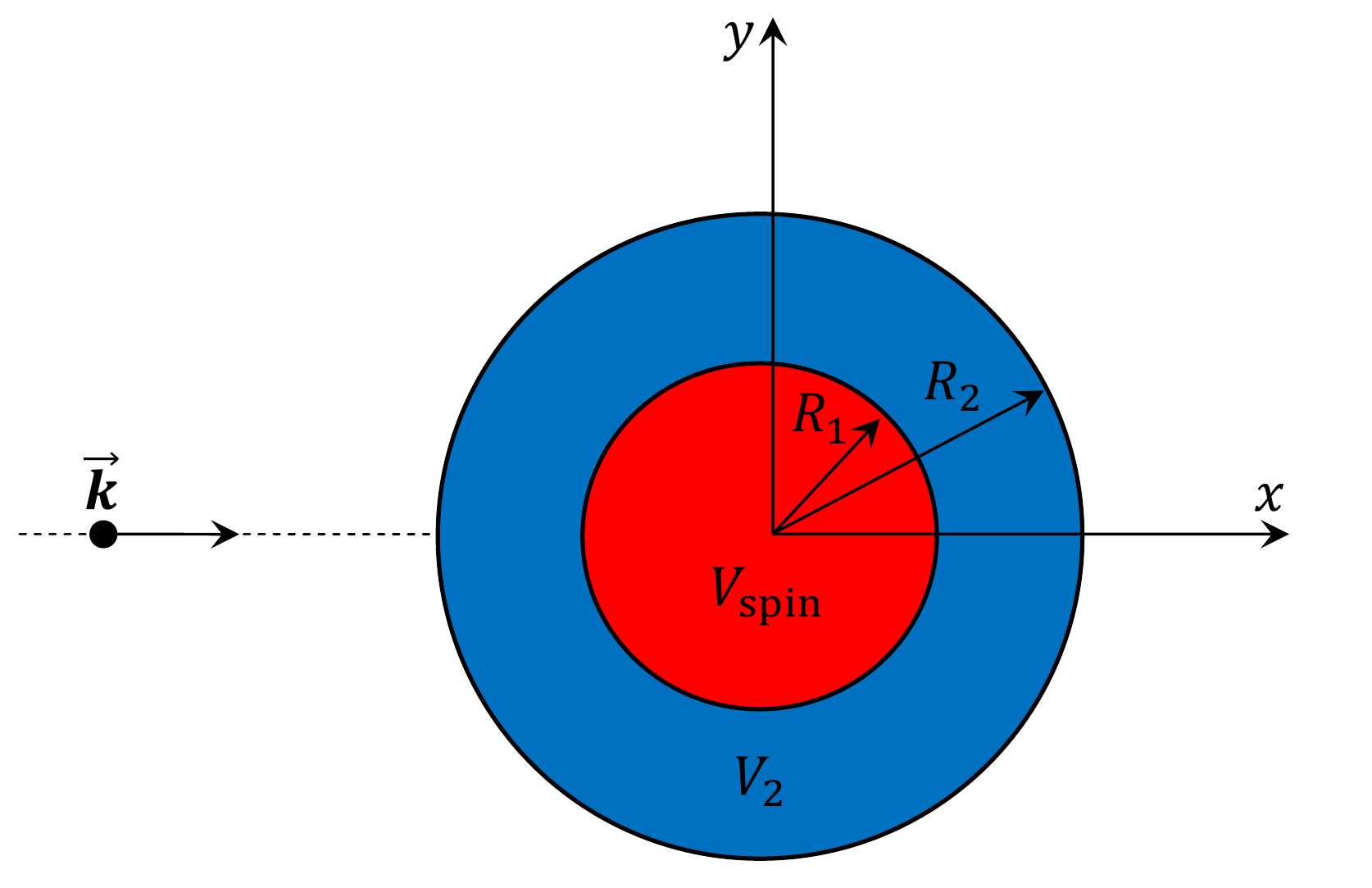}
\caption{Sketch of the cloaking setup, showing the incoming electron with a well defined energy $E$ and momentum $\vec{k}$, the impurity of radius $R_1$, and the cloak with internal radius $R_1$ and external radius $R_2$.}
\label{fig1}
\end{figure}

In this article we combine the idea of the electronic cloaking with the electronic properties of graphene to propose an alternative and efficient scheme for achieving the control of the spin scattering by an external agent.  For that purpose, we use an electron cloak similar to the ones proposed in References \onlinecite{Liao2012, Liao2013, Fleury2013, Oliver2015} (illustrated in Fig. \ref{fig1}) in which a a carrier with energy $E$ and well defined momentum is scattered off a radial core-shell. A gate surrounding the impurity (cloak) can induce cloaking effects by canceling a considerable part of the quantum scattering for a given energy window. We consider two different resonant scatterers: a magnetic impurity modeled by a local Zeeman potential along the $z$-direction and an adatom inducing spin-orbit coupling, modeled by a local intrinsic spin-orbit coupling. We show that we can change the strength and the sign of the spin polarization or the spin Hall angle by small variations of the gate voltage of the cloak where, in both cases, we achieve an efficient manipulation of the spin scattering for realistic values. Moreover, the achieved control of the SOC and ferromagnetism in graphene through proximity effect opens the possibility to fabricate large scattering centers and gates - in the order of hundreds of nanometers - and use the cloaking scheme for a precise control of the spin-dependent transport.

The article is organized as follows: in section \ref{sec2}, we introduce the spin dependence in the partial-waves formalism for graphene, and find the conditions for the resonant scattering produced by Zeeman-like and SOC impurities. In section \ref{sec3}, we analyze the effect of a gate voltage surrounding the two types of resonant scatterers, leading to the control of the spin dependent scattering by the cloaking scheme. Finally, in \ref{sec4} we summarize our results.

\section{Spin-dependent scattering}
\label{sec2}

We consider a single layer of graphene with an impurity in the shape of a disk that works as a scattering center. Inside the disk, there is a spin-dependent potential, to model either a magnetic moment or an impurity with SOC. This potential can be tuned to achieve the resonant scattering regime, where the peak of the transport cross-section occurs. Our starting point is the continuum-limit Hamiltonian of graphene

\begin{equation}
{\cal H}_{0}=\hbar v_{F}(\tau_{z}\sigma_{x}p_{x}+\sigma_{y}p_{y}) \text{,}
\label{eq2.1}
\end{equation}
where $\mathbf{p}=(p_{x},p_{y})$ is the momentum operator around one of the two nonequivalent Dirac points $K$ and $K^{\prime}$, $v_{F}\approx10^{6}$~m/s is the Fermi velocity, $\boldsymbol{\sigma}$ and $\boldsymbol{\tau}$ are Pauli matrices, with $\sigma_{z}=\pm1$ ($\tau_{z}=\pm1$) describing states on A-B sub-lattice (at $K$-$K^{\prime}$ Dirac points). Here, we consider scatterers large enough to neglect intervalley scattering. In addition, by taking the long wavelength limit, for potentials with radial symmetry, the disk scatterer is described by ${\cal H}_{V}={V}_{\text{spin}} \Theta(R-r)$ where $V_{\text{spin}}$ is a spin-dependent potential, $R$ is the radius of the scatterer and $\Theta(.)$ is the Heaviside function.

The Dirac equation for graphene carries an isospin associates to the sublattices. Therefore,  the spin dependence produces four-components eigenstates of the spin-dependent Hamiltonian for a single valley. To add the spin explicitly in the free-electron Hamiltonian, we choose the eigenstates of the $z$-component of the spin as basis and evaluate the tensorial product of the $2\times 2$ identity matrix $I$ with the Hamiltonian in \ref{eq2.1}: ${{\cal H}_{0}}_{\text{spin}} = I \otimes {\cal H}_{0} \text{.}$ The spin dependence leads to the eigenstates of ${{\cal H}_{0}}_{\text{spin}}$ that are two-component spinors for each spin component, resulting in a a four component wave-function $\Psi \left( \vec{r} \right)$.

Here we use the partial-waves method \cite{Peres2010,Novikov2007, Ferreira2011} to analyze the scattering process of a non-polarized electron beam by a radially symmetric potential that does not mix spin up and spin down states. In what follows, we derive the partial-wave scattering amplitudes and, from their knowledge, all gauge-invariant quantities can be determined unequivocally. In cylindrical coordinates, the four components of the graphene spinor $\Psi(\mathbf{r})=({\psi_{A}}^+({\mathbf{r}}),{\psi_{B}}^+({\mathbf{r}}),{\psi_{A}}^-({\mathbf{r}}),{\psi_{B}}^-({\mathbf{r}}))^{T}$ are decomposed in terms of radial harmonics

\begin{equation}
{\psi_{A}}^{\pm}({\mathbf{r}})=\sum_{m=-\infty}^{\infty} {g_{m}^{A}}^{\pm} (r)e^{im\theta}
\label{eq2.3}
\end{equation}
and

\begin{equation}
{\psi_{B}}^{\pm}({\mathbf{r}})=\sum_{m=-\infty}^{\infty} {g_{m}^{B}}^{\pm} (r)e^{i(m+1)\theta} \text{,}
\label{eq2.4}
\end{equation}
where $\theta\equiv\textrm{arg}(k_{x}^{\pm}+ik_{y}^{\pm})$, $\mathbf{k}$ is the wave vector, $m$ is the angular momentum quantum number, $A(B)$ represents the sub-lattice and $\pm$ represents the spin component $S_z$. After separating the variables of the graphene plus impurity Hamiltonian ${\cal H}_{0}+{\cal H}_{V_{\text{spin}}}$, we obtain four coupled first order equations for the radial functions ${g_{m}^{A}}^{\pm}(r)$ and ${g_{m}^{B}}^{\pm}(r)$, where $\boldsymbol{\mathbf{\sigma}}$ and $\mathbf{\boldsymbol{\tau}}$ are Pauli matrices for sub-lattice and valley respectively. As said before, $\tau_{z}$ is conserved and we can focus only on states at $K$ valley; scattering amplitudes for states at $K^{\prime}$ are quantitatively the same.

In a partial-waves expansion, the asymptotic form for the spinor wave-function for each spin component is given by \cite{Novikov2007, Peres2010}:

\begin{align}
\psi_{\lambda,\mathbf{k}}^{\pm}({\bf r}) & =\left(\begin{array}{c}
1\\
\lambda
\end{array}\right)e^{ikr\cos(\theta)}+\frac{f^{\pm}(\theta)}{\sqrt{-ir}}\left(\begin{array}{c}
1\\
\lambda e^{i{{\bf k}}}
\end{array}\right)e^{ikr} \text{,}
\label{eq2.5}
\end{align}
where $\lambda=\pm1$ brings the information about the carrier polarity, $f^{\pm}(\theta)$ is the spin-dependent scattering amplitude and the momentum $k=k^{\pm}$ also carries a spin dependence. By using a partial wave analysis, we can relate $f^{\pm}(\theta)$ with the phase-shifts $\delta_m^{\pm}$:

\begin{equation}
f^{\pm}(\theta)=\sqrt{\frac{2}{\pi k^{\pm}}}\sum_{m=-\infty}^{\infty}e^{im\theta}e^{i\delta_m^{\pm}}\sin{\delta_m^{\pm}}.
\label{eq2.6}
\end{equation}

Equation \ref{eq2.6} brings the possibility of different scattering amplitudes for up and down spin components, which results in a non-zero final polarization for $f_k^+ \left(\theta\right) \neq f_k^- \left(\theta\right)$, even with an initial non-polarized beam. We can evaluate the polarization of the scattered beam with:

\begin{equation}
P\left( \theta \right) = \frac{{\vert f^+\left( \theta \right) \vert}^2 - {\vert f^-\left( \theta \right) \vert}^2}{{\vert f^+\left( \theta \right) \vert}^2 + {\vert f^-\left( \theta \right) \vert}^2} \text{,}
\label{eq2.7}
\end{equation}
where the difference between the scattering amplitudes is normalized by their sum. In addition, we see an angular dependence in equation \ref{eq2.7}, which means that in principle we can have a different polarization according to the scattering angle.

Since we can analyze the scattering amplitudes separately for each spin component, it is intuitive to think that total differential cross-section is given by the sum of the spin up and spin down  contributions ${\frac{d\sigma}{d\theta}} \left( \theta \right) = {\frac{d\sigma}{d\theta}}^+ \left( \theta \right) + {\frac{d\sigma}{d\theta}}^- \left( \theta \right)$, where ${\frac{d\sigma}{d\theta}}^{\pm}=|f^{\pm}(\theta)|^2$. That been said, the longitudinal transport cross-section $\sigma_T(kR)$ and the transverse transport cross-section $\sigma_S(kR)$ can also be expressed separately for each spin component:

\begin{equation}
\sigma_T^{\pm}(kR)=\int {\frac{d\sigma}{d\theta}}^{\pm}(1-\cos(\theta))d\theta
\label{eq2.8}
\end{equation}
and
\begin{equation}
\sigma_S^{\pm}(kR)=\int {\frac{d\sigma}{d\theta}}^{\pm}\sin(\theta)d\theta \text{.}
\label{eq2.9}
\end{equation}

%

Similarly to the equations \ref{eq2.8} and \ref{eq2.9} we also can evaluate scattering parameters associated to the difference ${\vert f^+\left( \theta \right) \vert}^2 - {\vert f^-\left( \theta \right) \vert}^2$. They are the current spin polarizartion $P_S(kR)$ and the transport skewness $\gamma (kR)$, directly related to the spin Hall angle \cite{Ferreira2014}: 

\begin{equation}
P_S\left(kR\right) = \frac{\sigma_T^{+}(kR) - \sigma_T^{-}(kR)}{\sigma_T^{+}(kR) + \sigma_T^{-}(kR)}
\label{eq2.10}
\end{equation}
and
\begin{equation}
\gamma \left(kR\right) = \frac{\sigma_S^{+}(kR) - \sigma_S^{-}(kR)}{\sigma_T^{+}(kR) + \sigma_T^{-}(kR)} \text{.}
\label{eq2.11}
\end{equation}

\subsection{Magnetic impurities}
\label{subsec2.1}

We model the magnetic impurity as a local Zeeman potential along the $z$-direction, which allows us to define the scattering potential as follows:

\begin{equation}
V_{\text{spin}}\left(r\right)=
\left \{
\begin{array}{cc}
0, & r >a \\
V_z s_z, & r \leq a\\
\end{array}
\right. \text{.}
\label{eq2.12}
\end{equation}

The potential described in Eq. \ref{eq2.12} shows that for an incident electron (or hole) with spin up along the $z$-axis, the scatterer works as a potential barrier, while, for an incidence carrier with spin down, it works as a potential well. To identify the phase-shifts $\delta_m^{\pm}$, we write the spinors for the region inside and outside the potential as a superposition of angular harmonics. In the region $r>R$, we have $ k_{\text{out}} = \vert E \vert / \hbar v_{F} $ and the partial-wave $m$ is a sum of an incoming and a scattered wave according to 
\begin{align}
\psi_{m}^{\pm}(r,\theta) & =A_{m}^{\pm}\left(\begin{array}{c}
J_{m}(k_{\text{out}}r)e^{im\theta}\\
i\lambda_{\text{out}}J_{m+1}(k_{\text{out}}r)e^{i(m+1)\theta}
\end{array}\right)\nonumber\\&+B_{m}^{\pm}\left(\begin{array}{c}
Y_{m}(k_{\text{out}}r)e^{im\theta}\\
i\lambda_{\text{out}} Y_{m+1}(k_{\text{out}}r)e^{i(m+1)\theta}
\end{array}\right),
\label{eq2.13}
\end{align}
where $\lambda_{\text{out}} = sgn(E)$, whereas for $r<R$ we have just an incoming wave
\begin{equation}
\psi_{m}^{\pm}(r,\theta)=C_{m}^{\pm}\left(\begin{array}{c}
J_{m}(k_{\text{in}}^{\pm} r)e^{im\theta}\\
i\lambda_{\text{in}} J_{m+1}(k_{\text{in}}^{\pm} r)e^{i(m+1)\theta}
\end{array}\right),
\label{eq2.14}
\end{equation}
where $ k_{\text{in}}^{\pm} \equiv \vert E \mp V_z \vert / \hbar v_{F} $ and $\lambda_{\text{in}} = sgn(E \mp V_z)$. The continuity of the wave function at the interface of the potential leads to four equations, two for each spin component. Solving the two systems, we find the ratio $B_{m}^{\pm}/A_{m}^{\pm}$. It is straightforward to show that the phase-shift $\delta_{m}^{\pm}$ for partial-wave $m$ relates to the $B_{m}^{\pm}/A_{m}^{\pm}$ according to $B_{m}^{\pm}/A_{m}^{\pm}=-\tan(\delta_m^{\pm})$ (for more details about partial-waves in graphene, see references \cite{Novikov2007, Peres2010, Ferreira2011}).

The Dirac equation with Zeeman potential along the $z$-direction is symmetric under exchanging $g^{A \pm}_m$ and $ g^{B \pm}_{-(m+1)} $, which leads to the absence of back-scattering. The latter also corresponds to the relation $\delta_{m}^{\pm}=\delta_{-(m+1)}^{\pm}$ between phase-shifts, which leads to $\sigma_S^\pm = 0$ and, consequently, to $\gamma = 0$. Furthermore, for small energies $kR\ll1$, the channels $m=-1,0$ give the main contributions to $f^{\pm}(\theta)$ and, for a fixed energy  $E$ of the incoming electron, we can tune the potential $V$ to produce a resonant scatterer that will effectively trap the spin-polarized electrons inside the disk, which occurs for $\delta_{m}^{\pm}=\pm \pi/2$.

\begin{figure}[h]
\centering
\includegraphics[clip,width=\columnwidth]{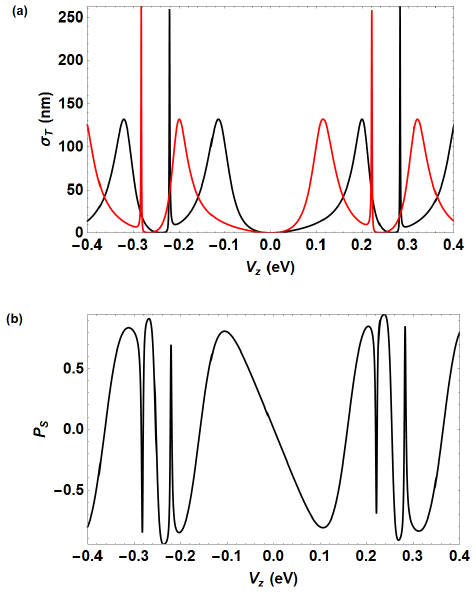}
\caption{(a) cross-section ${\sigma_T}_\downarrow$ (black) and ${\sigma_T}_\uparrow$ (red) and (b) current spin polarization $P_S$ (6 first partial waves are considered) as a function of the Zeeman potential $V_z$ for $E=0.02$ eV and $R=10$ nm.}
\label{fig2}
\end{figure}

The interaction of the incident beam with the Zeeman potential is different for each spin component, giving rise to two sets of resonances, one for each spin.  This is  reflected in the transport cross-section, where the resonances, that can be used to manipulate the charge transport in graphene \cite{Oliver2015}, are twice more often than in the case of a simple gated disk of a constant potential $V$. Figure \ref{fig2} (a) illustrates this feature: each one of the spin dependent transport cross-sections as a function of the potential present peaks that occur exactly at the maximum of the phase-shifts $\delta_m^\pm$, that characterize the resonant scattering regime. Figure \ref{fig2} (b) reveals that, for spin up peaks we always have positive current spin polarization, while for a resonance originated from the spin down scattering, we have $P_S<0$.

The setup presented in this section gives us a way to obtain highly spin polarizated scattered beams from an incoming electron with energy $E$  scattered by a Zeeman-like impurity with radius $R$ and potential $V_z$. This can be interpreted as a spin filter, since the combination of the three parameters can give rise to different  spin polarizations. It would be desirable to control $P_S$ by tuning the strength  $V_z$ of the Zeeman potential. However, from the experimental point of view this is quite challenging: the Zeeman potential is engineered either by a magnetic impurity or by proximity effect with a patterned magnetic insulator and in both cases its value is fixed. In addition, it is not possible to change the size of the magnetized region either. In section \ref{sec3} we discuss how to use the cloaking scheme to manipulate the spin polarization of the scattered beam by a gate.

\subsection{Impurity induced spin-orbit coupling}
\label{subsec2.2}

The large scatterers considered here induce a local SOC of the intrinsic type $V_{\text{so}}^{(I)} = \Delta_{\text{so}} \tau_z \sigma_z s_z$~\cite{Kane2005}; where $s_z$ is the Pauli matrix for the z-component of the spin. The impurity potential is assumed to be smooth on the lattice scale. For such large scatterers, inter-valley scattering is negligible ($ \tau_z = 1$) and, in the long wavelength limit, assuming that potentials have radial symmetry, the scatterer is described by:

\begin{equation}
V_{\text{spin}}\left(r\right)=
\left \{
\begin{array}{cc}
0, & r>a \\
V + \Delta_{\text{so}} \sigma_z s_z, & r \leq a\\
\end{array}
\right. \text{.}
\label{eq2.15}
\end{equation}

Where $V$ is scalar potential induced by the adatom or cluster \cite{Ferreira2014}. To identify $\delta_m^{\pm}$, we must write the spinors for the region outside (Equation \ref{eq2.13}) and inside the potential. For $r<a$, after solving the eigenvalue problem for ${\cal H} = {{\cal H}_{0}}_{\text{spin}} + V_{\text{spin}}$, we have $ k_{\text{in}} \equiv \sqrt{\epsilon^2 - \Delta_{\text{so}}^2} / \hbar v_{F}$, where $\epsilon \equiv E-V$. The wave functions are given by:
\begin{equation}
\psi_{m}^{\pm}(r,\theta)= \frac{C_{m}^{\pm}}{\sqrt{\epsilon}} \left(\begin{array}{c}
\sqrt{\epsilon \pm \Delta_{\text{so}}} J_{m}(k_{\text{in}}^{\pm} r)e^{im\theta}\\
i\lambda_{\text{in}} \sqrt{\epsilon \mp \Delta_{\text{so}}} J_{m+1}(k_{\text{in}}^{\pm} r)e^{i(m+1)\theta}
\end{array}\right),
\label{eq2.16}
\end{equation}
where $\lambda_\text{in} = sgn(\epsilon + \vert \Delta_{\text{so}} \vert)$. Applying the continuity of the wave function at the interface of the potential, we can find the ratio $B_{m}^{\pm}/A_{m}^{\pm}$ and, consequently, the phase-shifts $\delta_{m}^{\pm}$.

In the presence of intrinsic spin-orbit coupling, the Dirac equation is symmetric under exchanging ${g_{m}^{A}}^{\pm}$ and ${g_{-(m+1)}^{B}}^{\mp}$, which leads to the appearance of back-scattering. This symmetry also corresponds to the relation $\delta_{m}^{\pm}=\delta_{-(m+1)}^{\mp}$ between phase-shifts, which leads to $\sigma_T^\uparrow = \sigma_T^\downarrow (P_S =0)$ and $\sigma_S^+ = - \sigma_S^-$.  This corresponds to the generation of a spin current perpendicular to the initial beam, known as the spin Hall effect ($\gamma \neq 0$).

\begin{figure}[h]
\centering
\includegraphics[clip,width=\columnwidth]{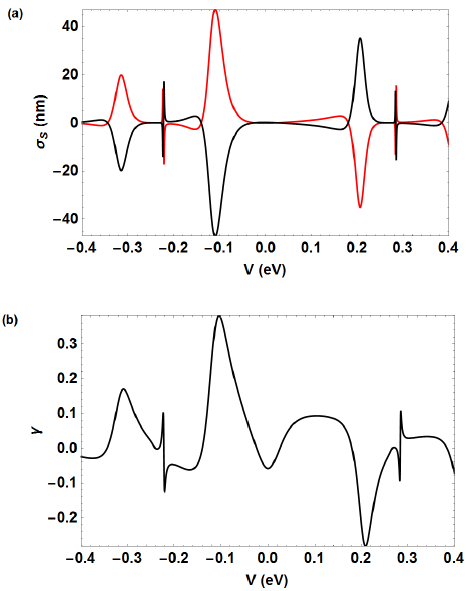}
\caption{(a) cross-section ${\sigma_S}_\downarrow$ (black) and ${\sigma_S}_\uparrow$ (red) and (b) transport skewness $\gamma$ (6 first partial waves are considered) as a function of the potential $V$ for $E=0.02$ eV, $R=10$ nm and ${\Delta}_{so}=25$ meV.}
\label{fig3}
\end{figure}

In Figure \ref{fig3}, we can see peaks at both skew cross-sections as a function of the potential. Again, the appearance of those peaks is a consequence of the phase-shifts $\delta_m^\pm$ characterizing resonant scattering. Figure \ref{fig3} (b) shows the transport skewness  $\gamma$, which also increases because of the resonance scattering. In addition, we can observe that the sign of $\gamma$ varies with the resonance and it is given by the symmetry or the spin-orbit coupling induced by the scatterer. From the experimental point of view, it is difficult to modify the characteristics of a impurity that produces spin-orbit coupling to manipulate the spin Hall effect. In the next section, we will see that a cloaking setup allows the control and enhancement of $\gamma$ by an external gate. 

\section{Spin dependent cloaking of resonant scatterers}
\label{sec3}

In general, a cloaking setup as illustrated in Fig. \ref{fig1} is used to produce invisibility, which in the case of electronic scattering, is equivalent to reducing the transport cross-section $\sigma_T$. Here, we use the cloak as an external agent,  to control the spin dependent scattering parameters, an consequently,  the spin polarization and the spin Hall angle. Our cloak is a homogeneous layer surrounding the impurity and can be implemented by using a gate, enabling the electric control of spin-dependent transport.

We begin by preparing the system in the resonant scattering regime, which maximizes the transport and skew cross-sections and the spin polarization. We now proceed to discuss the cloaking scheme. We keep the original potential $V_{\text{spin}}$ in the disk of radius $R$ (now called $R_1$) fixed, and  include a new potential $V_2$ in a ring of internal radius $R_1$, and external radius $R_2$ that is used to tune the scattering (see Figure \ref{fig1}).

We perform the same type of calculations described in the previous section but instead of defining the wave functions in two regions, we now have three. Outside the potentials (region 3), for $r>R_2$, we have
 
\begin{align}
{\psi_{m}^{3\pm}}(r,\theta) & =A_{m}^\pm\left(\begin{array}{c}
J_{m}(k_{3}r)e^{im\theta}\\
i\lambda_{3} J_{m+1}(k_{3}r)e^{i(m+1)\theta}
\end{array}\right)  \nonumber \\ 
& +B_{m}^\pm \left(\begin{array}{c}
Y_{m}(k_{3}r)e^{im\theta}\\
i\lambda_{3} Y_{m+1}(k_{3}r)e^{i(m+1)\theta}
\end{array}\right) \text{,}
\label{eq3.1}
\end{align}
with $k_3\equiv |E|/ \hbar v_{F}$ and $\lambda_3 = sgn(E)$, whereas for $R_2<r<R_1$ (region 2) we have 
\begin{align}
{\psi_{m}^{2\pm}}(r,\theta) & =C_m^\pm\left(\begin{array}{c}
J_{m}(k_{2}r)e^{im\theta}\\
i\lambda_{2} J_{m+1}(k_{2}r)e^{i(m+1)\theta}
\end{array}\right) \nonumber \\
&+D_{m}^\pm\left(\begin{array}{c}
Y_{m}(k_{2}r)e^{im\theta}\\
i\lambda_{2} Y_{m+1}(k_{2}r)e^{i(m+1)\theta}
\end{array}\right) \text{,}
\label{eq3.2}
\end{align}
where $k_2\equiv |E-V_2|/\hbar v_{F}$ and $\lambda_2 = sgn(E-V_2)$.

After writing the spinors of Dirac for the three regions, we apply the boundary conditions ${\psi_{m}^{3\pm}} (R_2,\theta)={\psi_{m}^{2}}^\pm(R_2,\theta)$ and ${\psi_{m}^{1 \pm}}(R_1,\theta)={\psi_{m}^{2 \pm}}(R_1,\theta)$, which leads to four equations, whose solution determine the ratio $B_{m}^\pm/A_{m}^\pm$ and, consequently, $\delta_m^\pm$. To highlight the effects of cloaking, we choose the parameters of $V_{\text{spin}}$  to achieve the resonant scattering regime. Then, we tune the potential of the ring $V_2$ and its radius $R_2$ in order manipulate the spin transport.

\subsection{Cloaking magnetic impurities}
\label{subsec3.1}

In the case of a magnetic impurity, we have the wave function 
\begin{equation}
{\psi_{m}^{1\pm}}(r,\theta)=E_{m}^\pm\left(\begin{array}{c}
J_{m}(k_{1}^\pm r)e^{im\theta}\\
i\lambda_{1} J_{m+1}(k_{1}^\pm r)e^{i(m+1)\theta}
\end{array}\right) \text{,}
\label{eq3.3}
\end{equation}
for $r<R_1$ (region 1) where $k_1\equiv |E \mp V_z|/\hbar v_{F}$ and $\lambda_1 = sgn(E \mp V_z)$.

After finding the phase-shifts $\delta_m^\pm$ by applying the boundary conditions and maximizing the effects of cloaking by putting the initial impurity in the resonant regime, we tune the potential of the ring $V_2$ and its radius $R_2$  to manipulate the value of the current spin polarizartion $P_S$.

Let us start with the situation described in the Figure \ref{fig1}, where we have a disk with radius $R_1=10\;\text{nm}$ and, for an incident beam with energy $E=0.02\;\text{eV}$ (a value close to the Dirac point, which can be easily accessible for graphene on h-BN), the resonant scattering occurs for some specific values of $V_z$. For $V_z=115\;\text{meV}$ and $V_z=199\;\text{meV}$, the broad peaks in the transport cross-section are related to the resonance of the low-order phase-shift, $\delta_0^-=\delta_{-1}^-$ and $\delta_0^+=\delta_{-1}^+$, respectively. On the other hand, for $V_z=221\;\text{meV}$ and $V_z=281\;\text{meV}$, we can see sharp peaks in consequence of the resonance of the phase-shifts $\delta_1^\pm=\delta_{-2}^\pm$, where  the lowest energy peak is the spin down resonance while the second is the spin up resonance. Next, we tune the potential and the radius of the ring to investigate the possibility of control of the current spin polarization $P_S$ by an external parameter. 

\begin{figure}[!ht]
\centering
\includegraphics[clip,width=\columnwidth,]{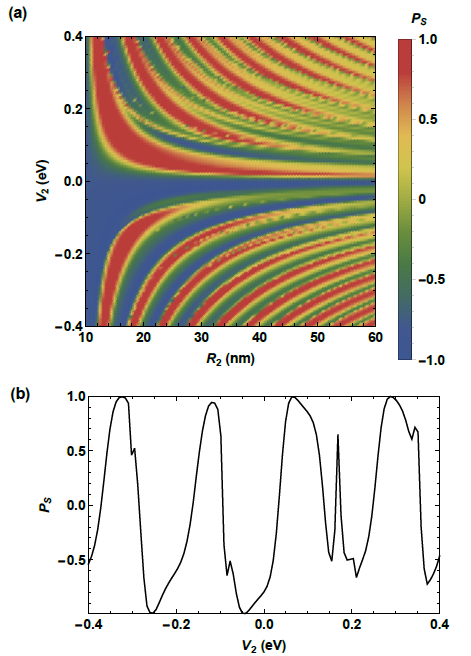}
\caption{(a)  Current spin polarization as a function of $V_2$ and $R_2$ for $E=0.02\;\text{eV}$, $R_1=10\;\text{nm}$ and $V_z=115\;\text{meV}$. (b) Spin polarization for the same parameters 
of (a) and $R_2=20$ nm.} 
\label{fig4}
\end{figure}

Figure \ref{fig4} (a) shows a density plot of the spin polarization as a function of the potential $V_2$  and radius $R_2$ of the ring, for a fixed value of $E$ and for the resonant scatterer with ${P_S}^0 = -0.788$ ($V_z=115\;\text{meV}$). Panel \ref{fig4} (b), which is a lateral cut of  Figure \ref{fig4} (a) for $R_2=20$ nm, reveals an very good  control of the spin polarization in function of $V_2$. We observe that it is possible to enhance the scattering rate for spin down ($P_S = -0.997$ for $V_2 = 46\;\text{meV}$) and  also suppress it, increasing the spin up scattering in a way that we get very close to $P_S = 1$. This  control of the spin polarization is possible for experimentally achievable  values for $R_2$ and $V_2$. In addition, for a fixed $R_2$, we get an oscillating picture of the transport spin polarization as a function of the the potential of the ring. These variations allow us to, with small variations of $V_2$, to easily change the sign or the amplitude of $P_S$ and go back to the initial value by continuos changes of the gate voltage.

\begin{figure}[h]
\centering
\includegraphics[clip,width=\columnwidth]{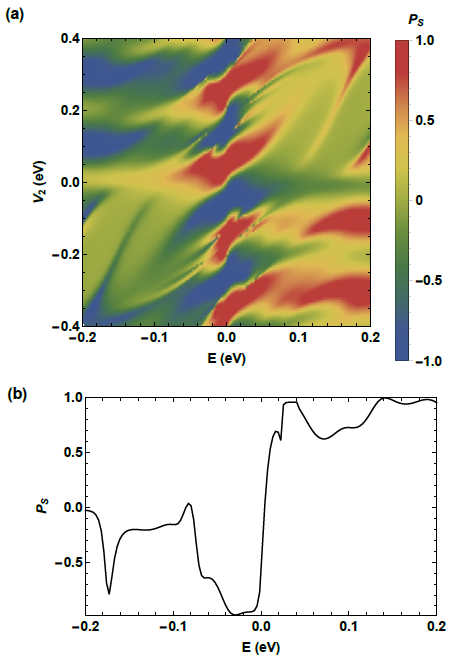}
\caption{(a) Spin polarization as a function of $V_2$ and $E$ for $R_1=10\;\text{nm}$, $V_z=115\;\text{meV}$ and $R_2=20\;\text{nm}$. (b) Spin polarization for the same parameters 
of (a) and $V_2=-300\;\text{meV}$.} 
\label{fig5}
\end{figure}

Figure \ref{fig5} (a) reports  $P_S$ as a function of the energy of the incident beam $E$ and $V_2$.  We can see that the maximum values for the transport spin polarization ($P_S = \pm 1$) can be obtained at low energies, close to the Dirac point, where the changes in the scattering regime are more pronounced. As we go far from $E=0$, one of the spin components has its scattering more enhanced, and we can achieve only one of the maximum spin polarizations for a given $E$. We can also set $V_2$ and analyze  the spin polarization in function of $E$.  Figure \ref{fig5} (b) highlights the energy selective nature of the spin scattering. 

\begin{figure}[h]
\centering
\includegraphics[clip,width=\columnwidth]{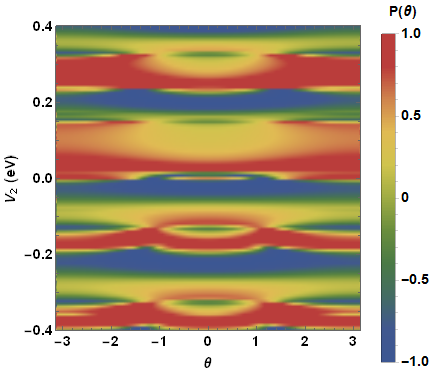}
\caption{Angular distribution of the spin polarization as a function of $V_2$ and the scattering angle $\theta$ for  $E=0.02\;\text{eV}$, $R_1=10\;\text{nm}$, $V_z=282\;\text{meV}$ and $R_2=20\;\text{nm}$}
\label{fig6}
\end{figure}

To obtain information on the direction of the scattering process, we mapped the angular distribution of the spin polarization a function of the scattering angle $\theta$ and $V_2$ (see Figure \ref{fig6}). We can see changes in the polarization according to the scattering angle, giving rise to a directional polarization, which consists of different intensities of the spin scattering for different angles of observation of the scattered beam. In addition, the tuning of $V_2$ allows the control of the scattering angle, demonstrating the feasibility of the directional scattering, since the directional polarization also gives which spin component has the higher scattering rate.  

In view of these results, we conclude that the incorporation of a gate as a cloak is a very efficient way of controlling the current spin polarization in this setup, consisting of a  magnetic adatom or cluster and tunable back and top gates \cite{Liao2012, Liao2013, Oliver2015}. In this particular case, to maximize the variations in the spin polarization, it is necessary to work with rings with radius in the order of tens of nanometers.

\subsection{Cloaking the SOC impurity}
\label{subsec3.2}

In the case of the impurity that generates the intrinsic spin-orbit, we have for $r<R_1$ (region 1) the wave function 
\begin{equation}
\psi_{m}^{\pm}(r,\theta)= \frac{E_{m}^{\pm}}{\sqrt{\epsilon}} \left(\begin{array}{c}
\sqrt{\epsilon \pm \Delta_{\text{so}}} J_{m}(k_{\text{1}}^{\pm} r)e^{im\theta}\\
i\lambda_{1} \sqrt{\epsilon \mp \Delta_{\text{so}}} J_{m+1}(k_{\text{1}}^{\pm} r)e^{i(m+1)\theta}
\end{array}\right)\text{,}
\label{eq3.4}
\end{equation}
where $k_{1} \equiv \sqrt{\epsilon^2 - \Delta_{\text{so}}^2} / \hbar v_{F}$ and $\lambda_\text{1} = sgn(\epsilon + \vert \Delta_{\text{so}} \vert)$. We use the same approach of Section \ref{subsec3.1}:  we apply the boundary conditions to find the phase-shifts $\delta_m^\pm$ and set the initial impurity in the resonant scattering regime. The major difference here is that we tune the potential of the ring $V_2$ and its radius $R_2$ to manipulate the value of the transport skewness $\gamma$.

Beginning with the scheme presented in the Figure \ref{fig1}, we set $R_1=10\;\text{nm}$ and $E=0.02\;\text{eV}$. In the following, by looking at the Figure \ref{fig3}, we know the values of $V$ that will providuce the resonant regime for the initial impurity. For $V=202\;\text{meV}$ and $V=-116\;\text{meV}$, we have large peaks at the skew cross-section because of the resonance of the phase-shifts $\delta_0^{\pm}=\delta_{-1}^\mp$, and for $V=284\;\text{meV}$ and $V=-224\;\text{meV}$, we observed sharp peaks due to the next phase-shifts in order of contribution, $\delta_1^{\pm}=\delta_{-2}^\mp$. Next, we tune the potential and the radius of the ring to investigate the control of $\gamma$ in these particular situations. To analyze the efficiency of using the spin cloaking to maximize the spin Hall effect, we look at the ratio $\gamma/\gamma_0$ between the skewness with and without the ring - $\gamma$, and $\gamma_{0}$ respectively.

\begin{figure}[h]
\centering
\includegraphics[clip,width=\columnwidth]{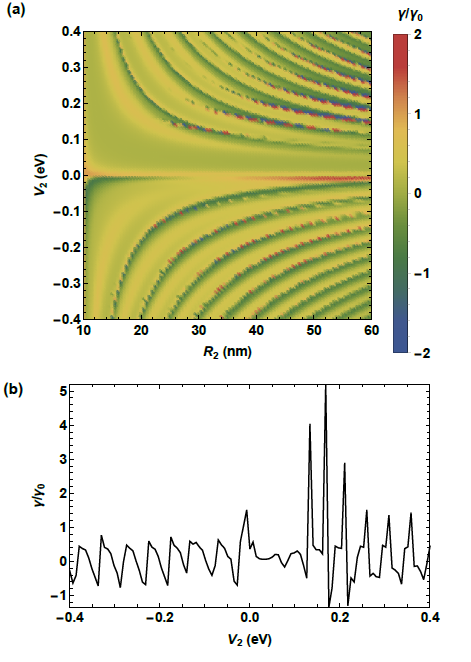}
\caption{(a) Spin cloaking efficiency as a function of $V_2$ and $R_2$ for $E=0.02\;\text{eV}$, $R_1=10\;\text{nm}$, $V=284\;\text{meV}$ and $\Delta_{\text{so}} = 25\;\text{meV}$. (b) Spin cloaking efficiency for the same parameters of (a) and $R_2=50\;\text{nm}$.} 
\label{fig7}
\end{figure}

Figure \ref{fig7}(a) shows a density plot of the spin cloaking efficiency as a function of the potential $V_2$ and radius $R_2$ of the ring, for a fixed value of $E$ and of the spin Hall angle for the initial resonant scatterer $\gamma_{0}= 0.105$ ($V=284\;\text{meV}$). This plot reveals that, for the potential of the ring working as cloak, we can enhance the initial spin Hall angle and invert its signal for realistic values of $V_2$ and $R_2$. Besides that, Figure \ref{fig7} (b) shows us that, for a fixed value of the external ratio ($R_2=50\;\text{nm}$), we have an oscillating picture where we can invert the spin current just by tuning $V_2$ in a few meV, which allows us to choose the spin scattering scenario by small variations under $V_2$.

\begin{figure}[h]
\centering
\includegraphics[clip,width=\columnwidth]{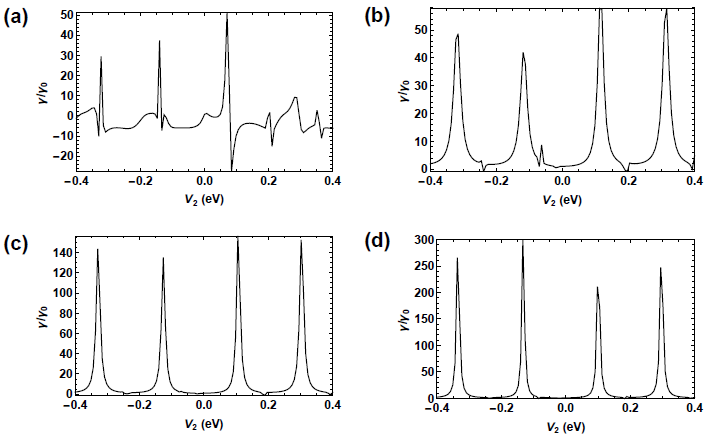}
\caption{(a) Spin cloaking efficiency as a function of $V_2$ for $R_1=10\;\text{nm}$, $V=284\;\text{meV}$, $\Delta_{\text{so}} = 25\;\text{meV}$ and $R_2=20\;\text{nm}$ for different energies: (a) $E=0.04\;\text{eV}$, (b) $E=0.004\;\text{eV}$, (c) $E=0.002\;\text{eV}$ and (d) $E=0.001\;\text{eV}$.} 
\label{fig8}
\end{figure}

Following the same analysis of the Zeeman impurities, we now check how the spin cloaking efficiency behaves for variations of the energy of the initial beam. We see that the enhancement of the initial spin Hall angle is larger as we get close to $E=0$. For example, for $E=0.001\;\text{eV}$ (Figure \ref{fig8} (d)) we can observe a skew polarization 300 times higher than the initial one, as for $E=0.002\;\text{eV}$ (Figure \ref{fig8} (c)) we can observe its enhancement in 140 times. On the other had, for higher energies ($E=0.04\;\text{eV}$, for example), we can get an inversion of the initial $\gamma_{0}$, as seen in Figure \ref{fig8} (a).

\section{Conclusions}
\label{sec4}

In conclusion, we have observed a good control of the spin-dependent electronic scattering by a cloaking mechanism in graphene. This mechanism consists of a gate in the shape of a ring that works as a cloak around a scattering center. In general, cloak can be tuned to reduce the electronic scattering cross-section but we have shown here that is can also be used to manipulate the current spin polarization and the spin Hall angle in graphene with spin-dependent impurities. Here, we consider two types of spin-dependent scattering centers: Zeeman-like and intrinsic spin-orbit impurities with a spin independent cloak. We use a partial-waves expansion with explicit spin dependence and demonstrate that this electronic cloaking mechanism is very efficient: we observe a very good control of the spin parameters of the scattering as a function of an external agent, represented here by the voltage of the cloak. For the Zeeman-like impurity, it is shown that we can enhance the initial spin polarization or change its signal by small changes of the potential of cloak, while, for the SOI impurity, we observe the same control, but for the spin Hall angle, that is related to the generation of a perpendicular spin current due to the spin Hall effect. Our results suggest that this setup could be explored in spintronics, more precisely in applications aiming at tuning the spin flow in graphene and in spin-based devices.

\section*{Acknowledgements}

We acknowledge the Brazilian agencies CNPq, CAPES and FAPERJ for financial support. T. G. Rappoport  acknowledges the financial support of the Royal Society (U.K.) through a Newton Advanced Fellowship and thanks the University of York for the administrative support.

\end{document}